# Superconducting properties of $BiS_2$ based superconductor $NdO_{1-x}F_xBiS_2$ ($x$=0 to 0.9)

Rajveer Jha, and V.P.S. Awana

CSIR-National Physical Laboratory, Dr. K.S. Krishnan Marg, New Delhi-110012, India

**Abstract**: We report the phase diagram of $NdO_{1-x}F_xBiS_2$ ($x$=0.1, 0.3, 0.5, 0.7, 0.9) compound. The samples synthesized by the vacuum encapsulation technique are crystallized in a tetragonal $P4/nmm$ space group. The superconductivity of all samples is confirmed by *DC* magnetic susceptibility and electrical transport measurements. The highest superconducting transition temperature ($T_c$) onset of 5.25K is seen for $NdO_{0.3}F_{0.7}BiS_2$ sample. Superconductivity is suppressed with applied magnetic field, as both $T_c^{onset}$ and $T_c$ ($\rho$ =0) are decreased to lower temperatures in magneto transport $\rho(T, H)$ measurements. Upper critical field $H_{c2}(0)$ is estimated to be above 3Tesla for $NdO_{0.3}F_{0.7}BiS_2$ sample. The Arrhenius equation is used to calculate the thermally activated flux flow (*TAFF*) activation energy ($U_0$) for $NdO_{0.3}F_{0.7}BiS_2$ sample, which is 61.64 meV in 0.02Tesla field. Hall Effect measurements show that the electron charge carriers are the dominating ones through out the studied temperature range of 300K – 2K.

*Corresponding Author
Dr. V. P. S. Awana, Principal Scientist
E-mail: awana@mail.npindia.org
Ph. +91-11-45609357, Fax-+91-11-45609310
Homepage www.fteewebs.com/vpsawana/

**Introduction:**

Recently discovered layered $BiS_2$ based superconductors, with $T_c$=8.6K in $Bi_4O_4S_3$ and $T_c$=3-10 K in (La/Nd/Ce/Pr)$O_{0.5}F_{0.5}BiS_2$ have been reported by many scientific groups [1-10]. Successful growth of the superconducting $ReO_{1-x}F_xBiS_2$ (Re-La, Nd, & Ce) single crystals confirmed the intrinsic nature of superconductivity in these materials [11, 12]. The $BiS_2$ layer of these layered compounds is responsible for superconductivity, which is similar to the $CuO_2$ planes in high $T_c$ cuprate superconductors [13] and the FeAs planes in iron based superconductors [14]. The parent phase of $ReOBiS_2$ (Re= La, Nd, Ce, & Pr) compounds exhibits bad metals or a Mott insulator behavior [5, 15]. The substitution of $F^{-1}$ at $O^{-2}$ in the oxide blocks (ReO) induces superconductivity in these compounds via electron doping. The ReO layer acts as spacer block in $ReO_{1-x}F_xBiS_2$ compounds. Superconductivity has also been observed via hole doping i.e., by $La^{3+}$ substitution at the site of $Sr^{2+}$ site in similar structure $SrFBiS_2$ compound [16]. More recently, observation of superconductivity is reported via substitution of tetravalent $Th^{+4}$, $Hf^{+4}$, $Zr^{+4}$, and $Ti^{+4}$ for trivalent $La^{+3}$ in $LaOBiS_2$ compound [17]. Basically superconductivity appears by various carrier doping routes in the vicinity of the insulating-like state of the un doped parent $BiS_2$-based systems viz. $ReOBiS_2$ and $SrFBiS_2$ [5, 15-17]. Thus these systems are very sensitive to the carrier doping level, and superconducting properties can be changed with the variation of doping concentration level [9, 16]. The size of the band gap between the conduction and the valence bands play an important role in determining the actual carrier concentration. As expected, electrons are significantly doped from the nominal fluorine content in $NdO_{1-x}F_xBiS_2$ [9, 18]. Hall Effect measurements suggest that though electron are the dominant carriers, the multiband features are also seen and the superconducting pairing occurs in the quasi two dimensional chain in these systems [15]. Theoretical studies show that the electron carriers are originated from the Bi 6p orbital and the superconductivity occurs in the $BiS_2$ plane [18, 19]. Thus the fundamental parameters, such as the upper critical field $H_{c2}$, charge carrier density and their type etc, are important to identify the superconducting mechanism in these novel $BiS_2$ based superconductors. It is well known that the normal state properties, such as the Hall Effect and magneto-resistance, can lend crucial information about the fundamental parameters of a material.

In this paper, we extend our previous work on $NdO_{0.5}F_{0.5}BiS_2$ [20] superconductor to see the effect of Fluorine composition in $NdO_{1-x}F_xBiS_2$ (x=0.1, 0.3, 0.5, 0.7 & 0.9) compounds on superconducting transition temperature ($T_c$) and other fundamental parameters. All the synthesized samples are crystallized in tetragonal *P*4/*nmm* space group. With increasing the doping level of F the lattice parameter (*c*) decreases till *x*=0.7. The superconducting temperature with $T_c^{onset}$ ~5.25K is highest for nominal Fluorine doping of x=0.7 and the same is decreased for higher concentration of x=0.9. We present the AC susceptibility, electrical resistivity under magnetic field and Hall Effect measurement of the maximum $T_c$ $NdO_{0.3}F_{0.7}BiS_2$ sample.

**Experimental:**

The polycrystalline bulk $NdO_{1-x}F_xBiS_2$ (x=0.1, 0.3, 0.5, 0.7 & 0.9) samples are synthesized by standard solid state reaction route via vacuum encapsulation. High purity Nd,

Bi, S, $NdF_3$, and $Nd_2O_3$ are weighed in stoichiometric ratio and ground thoroughly in a glove box in high purity argon atmosphere. The mixed powders are subsequently palletized and vacuum-sealed ($10^{-4}$ mbar) in a quartz tube. The box furnace have been used to sinter the samples at $780^0C$ for 12h with the typical heating rate of $2^oC/min$. and sintered samples subsequently cooled down slowly to room temperature. This process was repeated twice. X-ray diffraction (*XRD*) was performed at room temperature in the scattering angular ($2\theta$) range of $10^o$-$80^o$ in equal $2\theta$ step of $0.02^o$ using *Rigaku Diffractometer* with *Cu $K_\alpha$* ($\lambda = 1.54$Å). Rietveld analysis was performed using the standard *FullProf* program. Detailed DC & AC transport and magnetization measurements were performed on Physical Property Measurements System (*PPMS*-14T, *Quantum Design*).

**Results and discussion:**

The room temperature observed and Reitveld fitted XRD pattern of as synthesized $NdO_{1-x}F_xBiS_2$ (x=0.1-0.9) samples are shown in Figure 1. All compounds are crystallized in tetragonal structure with space group *P4/nmm*. Small impurity peaks of NdOF and $Bi_2S_3$ are also been observed, which are marked in Figure 1. It can be seen from the XRD pattern the (004) shifts towards the higher $2\theta$ angle, suggesting the smaller *c*-axis constant. Reitveld analysis of XRD show that the lattice parameter *c* decreases from 13.509Å to 13.404Å as *x* is increased from $x = 0.1$ to 0.7 and the same again increases (*c*=13.406Å) for x=0.9. It seems the $x = 0.7$ is the solubility limit of F at O site in $NdO_{1-x}F_xBiS_2$. Decrease in *c* parameter is clear indication of the fact, that the layer structure shrinks with successive $O^{-2}$ site $F^{-1}$ substitution in the $NdO_{1-x}F_xBiS_2$ system. These results are in good agreement with the reported phase diagrams for $NdO_{1-x}F_xBiS_2$, $CeO_{1-x}F_xBiS_2$ and $PrO_{1-x}F_xBiS_2$ compound [9, 15, 21].

Figure 2 depict the DC magnetic susceptibility being performed in both ZFC (Zero Field Cool) and FC (Field Cool) protocols under the applied magnetic field 10Oe for the $NdO_{1-x}F_xBiS_2$ (x=0.1, 0.3, 0.5, 0.7) samples. Superconducting transitions for these measured samples are observed in terms of clear diamagnetic signals below their respective $T_c$. The strongest diamagnetic signal $4\pi\chi$ in ZFC at 2K is near −1 for the $NdO_{0.3}F_{0.7}BiS_2$ sample. This shows that the shielding volume fraction for $NdO_{0.3}F_{0.7}BiS_2$ sample at 2K is almost 100%. The diamagnetic signal is also observed at below 2.3K for $x = 0.1$, 3.5K for $x = 0.3$, and 4.6K for x=0.5 samples with their respective shielding volume fractions of around 5%, 25%, and 50% at 2K. The $NdO_{0.3}F_{0.7}BiS_2$ sample exhibits the maximum shielding volume fraction (~100%) with highest $T_c$ of above 5K. The *ac* magnetization of maximum $T_c$ sample $NdO_{0.3}F_{0.7}BiS_2$ being measured at different ac field amplitudes (0.5-15 Oe) and fixed frequency of 333Hz with zero bias dc field is shown in Fig. 3. The *ac* magnetic susceptibility, confirms the bulk superconductivity at around 5K. It can be seen, that the height of imaginary part peak as well as diamagnetism in real part of ac susceptibility are increased monotonically with increase in amplitude. On the other hand the imaginary part peak position temperature (4K) is not changed with increasing *ac* amplitude, suggesting that the superconducting grains are well coupled in $NdO_{0.3}F_{0.7}BiS_2$ superconductor.

The normalized resistance ($R/R_{300}$) vs temperature is shown in the Fig. 4 for the $NdO_{1-x}F_xBiS_2$ (x=0, 0.1, 0.3, 0.5, 0.7, 0.9) samples. The parent phase of $NdO_{1-x}F_xBiS_2$ compound shows semimetal behavior in the temperature range of 300-2K. With successive doping of $F^{-1}$ at $O^{-2}$ the superconductivity starts appearing in $NdO_{1-x}F_xBiS_2$ compounds. The highest $T_c$ above 5.25K is seen at the nominal concentration of x =0.7, and later $T_c$ decrease for x=0.9. A superconducting onset transition is observed for all the F doped samples but for x=0.9, $T_c(R=0)$ is not seen down to 2K. Inset of the Fig.4 shows the extended plot of $R/R_{300}$ vs $T$ from 2- 6K. it is clear that all the F doped samples show the $T_c^{onset}$, namely at 2.5K, 3.5K, 4K, 4.8K and 5.25K for x=0.1, 0.3, 0.9, 0.5 and 0.7 respectively. Figure 5 shows the nominal *x* dependence of lattice parameter *c* (Å), and $T_c^{onset}$. It is clear that $T_c$ of $NdO_{1-x}F_xBiS_2$ increases with nominal Fluorine concentration (x) until x=0.7 along with a decrease in *c* lattice parameter. One can say the $T_c$ is highest for lowest *c* lattice parameter, probably as besides the carrier doping the chemical pressure is increasing with the concentration of the *x* till x = 0.7. For x= 0.9 sample, the lattice constant is increased and the $T_c^{onset}$ is decreased. Worth mentioning is the fact that $T_c^{onset}$ values being seen from resistivity measurements (Fig.4) are slightly higher than the ones being determined from the magnetization measurements (Fig. 2 and 3), this is precisely because the former is percolation path for current flow and the later is bulk property of a superconductor.

Figure 6 (a), (b), and (c) show the temperature dependence of resistivity under applied magnetic field for three samples with x=0.3, 0.5, and 0.7 respectively. All the three samples shows relatively sharp superconducting transitions at $T_c$ = 3.5K, 5.0K and 5.25K. With applied field, one can see that both the $T_c^{onset}$ and $T_c(\rho=0)$ shift towards the low temperature. The former one is controlled by the upper critical field of the individual grains in a superconductor, while the latter one is due to the weak links between the grains as well as the vortex flow behavior. To determine upper critical field, we have taken 90% of normalized resistivity of the onset point of the transition and the conventional one-band Werthamer–Helfand–Hohenberg (*WHH*) equation  ($H_{c2}(0)=−0.693T_c(dH_{c2}/dT)_{T=Tc}$) is used to determine the  upper critical field [22]. Figure 7 shows the $H_{c2}$ versus $T$ for $NdO_{0.7}F_{0.3}BiS_2$, $NdO_{0.5}F_{0.5}BiS_2$ and $NdO_{0.3}F_{0.7}BiS_2$ samples. The solid lines represent the fitted $H_{c2}(T)$ from *WHH* formula with respect to the experimentally observed data points. The values of $H_{c2}^{WHH}(0)$ are 0.9 T for $NdO_{0.7}F_{0.3}BiS_2$, 2.8T for $NdO_{0.5}F_{0.5}BiS_2$, and 3.3T for $NdO_{0.3}F_{0.7}BiS_2$.

The plots of temperature derivative of resistivity for the highest $T_c$ compound $NdO_{0.3}F_{0.7}BiS_2$ at various applied magnetic field are shown in Fig. 8. It is clear from Fig. 8 that the broadening of $d\rho/dT$ peak takes place with increase in applied magnetic field, which suggests that the superconducting onset is relatively affected less than the $T_c$ ($\rho$=0) state. The temperature derivative of resistivity in superconducting transition region gives narrow intense maxima, centered at mid of $T_c$ in zero field, indicating good percolation path of superconducting grains. In this plot we can clearly see single transition peak at every applied field, which is an indication towards the better grains coupling in this compound. The broadening of the resistive transitions in the $\rho(T)H$ curve with increasing magnetic fields can be seen in terms of energy dissipation due to vortex motion [23]. The resistivity in broadened region is caused by the creep of vortices, which are thermally activated. The resistivity in *TAFF* regime of the flux creep is given by Arrhenius equation [24],

$$\rho(T,B) = \rho_0 \exp[-U_0/k_B T]$$

Where, $\rho_0$ is the field independent pre-exponential factor (here normal state resistance at 6K $\rho_{6K}$ is taken as $\rho_0$), $k_B$ is the Boltzmann's constant and $U_0$ is *TAFF* activation energy. To reveal the origin of thermally activated $\rho(T,B)$, we present the Arrhenius plots of representative low resistance data. The linear dependence of ln $\rho$ vs. $T^{-1}$ plot for $NdO_{0.3}F_{0.7}BiS_2$ sample is shown in Fig. 9. The values of the activation energy in the magnetic field range 0.02 to 2T are estimated from 62.66meV to 0.12meV.

Figure 10 represent the Hall Effect measurements of superconducting $NdO_{0.3}F_{0.7}BiS_2$ sample. The Hall Effect measurement is performed to demonstrate the strange normal state behavior and obtain some useful properties such as suggested charge density wave instability in this compound [25]. The Hall coefficient ($R_H$) of $NdO_{0.3}F_{0.7}BiS_2$ sample is measured under 1T applied magnetic field in the temperature range 2 to 300K. We observed the negative $R_H$ signal for the given temperature range, which suggest that the electrons are the dominating charge carriers in this system, same as for Fe based superconductors [26]. Inset of the Fig 10 shows magnetic field dependence of Hall resistivity ($\rho_{xy}$), which is measured with magnetic field being perpendicular to the current and surface of the sample. The voltage $V_{xy}$ is recorded across the direction of sample width. The sign of Hall resistivity $\rho_{xy}$ at various temperatures of 2K, 10K, 50K, 100K, 200K and 300K of the studied $NdO_{0.3}F_{0.7}BiS_2$ sample remains negative. Here we can see that the field dependence $\rho_{xy}$ exhibits curved feature at 2K and for higher temperatures resistivity is linearly field dependent and remains negative. The curved feature at 2K happens due to the fact that the compound is superconducting at this temperature below its upper critical field. The single-band assumption is used to estimate the charge-carrier densit. As the $\rho_{xy}$ exhibits linear behavior with the magnetic field, the $R_H$ can be measured by $R_H = d\rho_{xy}/dH = 1/ne$, for a single band metal, where $n$ is the charge carrier density. For the single band assumption we estimate the charge carrier density of about 1.278 × $10^{19}$/cm$^3$ at 10K.

**Conclusion:**

In conclusion, we have successfully synthesized BiS$_2$-based superconductor NdO$_{1-x}$F$_x$BiS$_2$ (x=0.1-0.9) by solid state vacuum encapsulation technique. Reitveld refinement of XRD pattern show that all the samples are crystallized in a tetragonal *P4/nmm* space group and the lattice parameters decrease with the F-doping. The bulk superconductivity with maximum $T_c$ of 5.25K and nearly 100% shielding volume fraction at 2K is seen for nominal *x*=0.7. Electrical resistivity measurements showed that under applied magnetic field both $T_c$ onset and $T_c$ ($\rho$ =0) decrease to lower temperatures and an upper critical field $H_{c2}(0)$ of above 3Tesla is estimated for NdO$_{0.3}$F$_{0.7}$BiS$_2$ sample. The activation energy in the magnetic field range 0.02 to 2Tesla has been estimated from 62.66meV to 0.12meV. Hall measurement results indicated dominance of electron charge carriers in this compound.

**Acknowledgement:**
Authors would thank their Director NPL, India for his consistent support and interest in the present work. This work is financially supported by DAE-SRC outstanding investigator award scheme on search for new superconductors. Rajveer Jha acknowledges the CSIR for the senior research fellowship.**Reference:**
1. Mizuguchi Y, Fujihisa H, Gotoh Y, Suzuki K, Usui H, Kuroki K, Demura S, Takano Y, Izawa H, Miura O, 2012 Phys. Rev. B **86**, 220510(R).

2. Singh S K, Kumar A, Gahtori B, Shruti, Sharma G, Patnaik S, Awana V P S, 2012 J. Am. Chem. Soc. **134**, 16504.

**Figure Captions**

**Figure 1:** Observed (*open circles*) and calculated (*solid lines*) XRD patterns of $NdO_{1-x}F_xBiS_2$ (x=0.1, 0.3, 0.5, 0.7 and 0.9) compound at room temperature.

**Figure 2:** *DC* magnetization (both *ZFC* and *FC*) plots for $NdO_{1-x}F_xBiS_2$ (x=0.1, 0.3, 0.5 & 0.7) measure in the applied magnetic field, H= 10 Oe.

**Figure 3:** The ac magnetic susceptibility in real (*M'*) and imaginary(*M''*) situations at fixed frequency of 333 Hz in varying amplitudes of 0.5–15 Oe for $NdO_{0.3}F_{0.7}BiS_2$ sample.

**Figure 4:** Resistance versus temperature (*$R/R_{300}$* Vs *T*) plots for $NdO_{1-x}F_xBiS_2$ (x=0, 0.1, 0.3, 0.5, 0.7 & 0.9) samples, inset show the same in 2-6.0 K temperature range.

**Figure 5:** The *nominal x* dependence to the lattice constants (*c*-axis) and superconducting $T_c$ is shown in the plot.

**Figure 6:** (a), (b) & (c) shows Temperature dependence of the resistivity ρ(T) under magnetic fields for the samples $NdO_{0.7}F_{0.3}BiS_2$, $NdO_{0.5}F_{0.5}BiS_2$ and $NdO_{0.3}F_{0.7}BiS_2$ respectively.

**Figure 7:** The upper critical field $H_{c2}$ found from 90% of the resistivity ρ(T) for the samples $NdO_{0.7}F_{0.3}BiS_2$, $NdO_{0.5}F_{0.5}BiS_2$ and $NdO_{0.3}F_{0.7}BiS_2$.

**Figure 8:** Temperature derivative of normalized resistivity *Vs T* of $NdO_{0.3}F_{0.7}BiS_2$ sample.

**Figure 9:** Fitted Arrhenius plot of resistivity for $NdO_{0.3}F_{0.7}BiS_2$ samples.

**Figure 10:** The Hall resistivity $\rho_{xy}$ vs the magnetic field $\mu_0H$ at 2, 10, 50, 100, 200, and 300K of the sample $NdO_{0.3}F_{0.7}BiS_2$. Inset shows the Hall coefficient $R_H$ of the same sample at 1T from 2 to 300K.

**Figure 1**

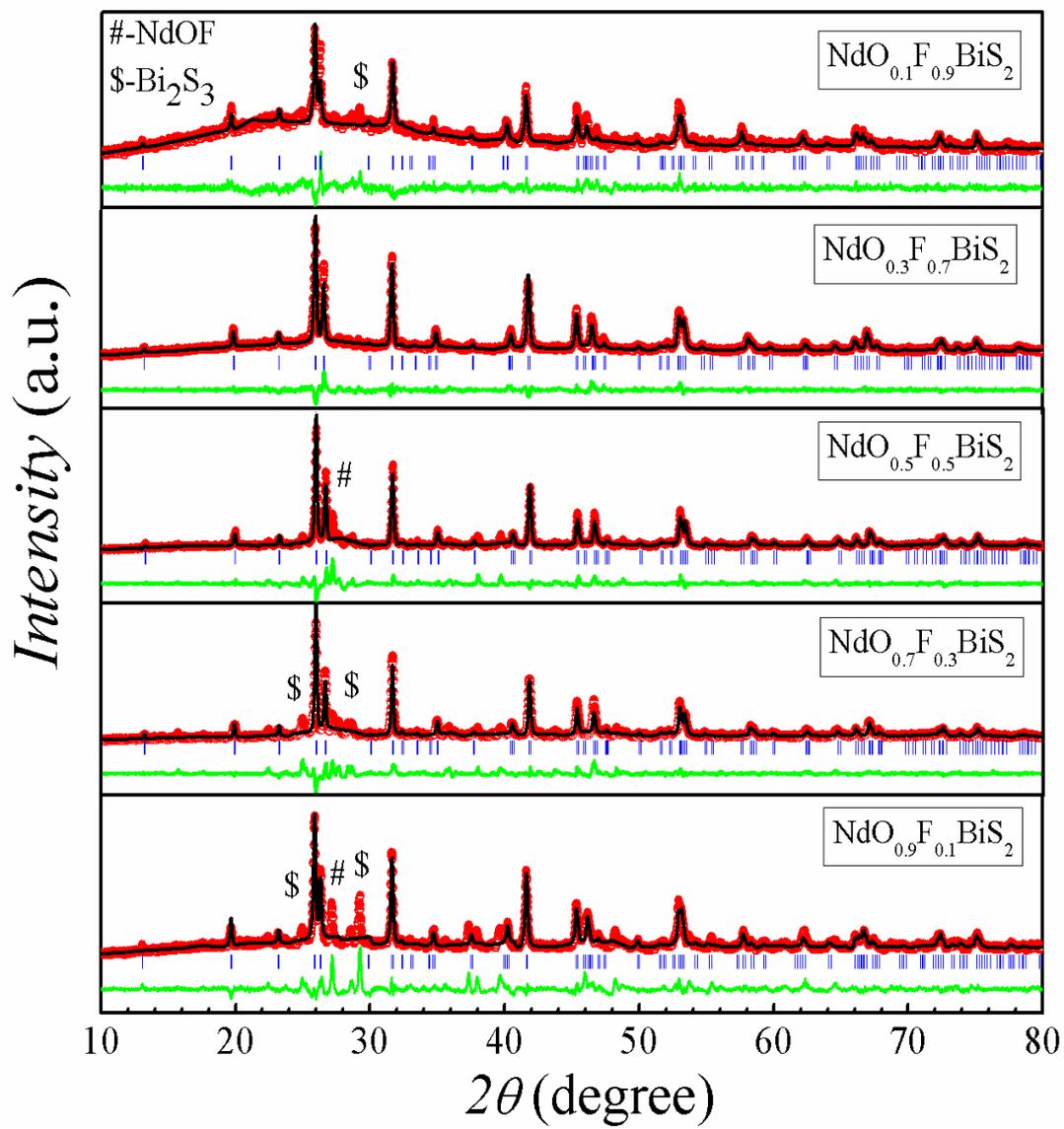

**Figure 2**

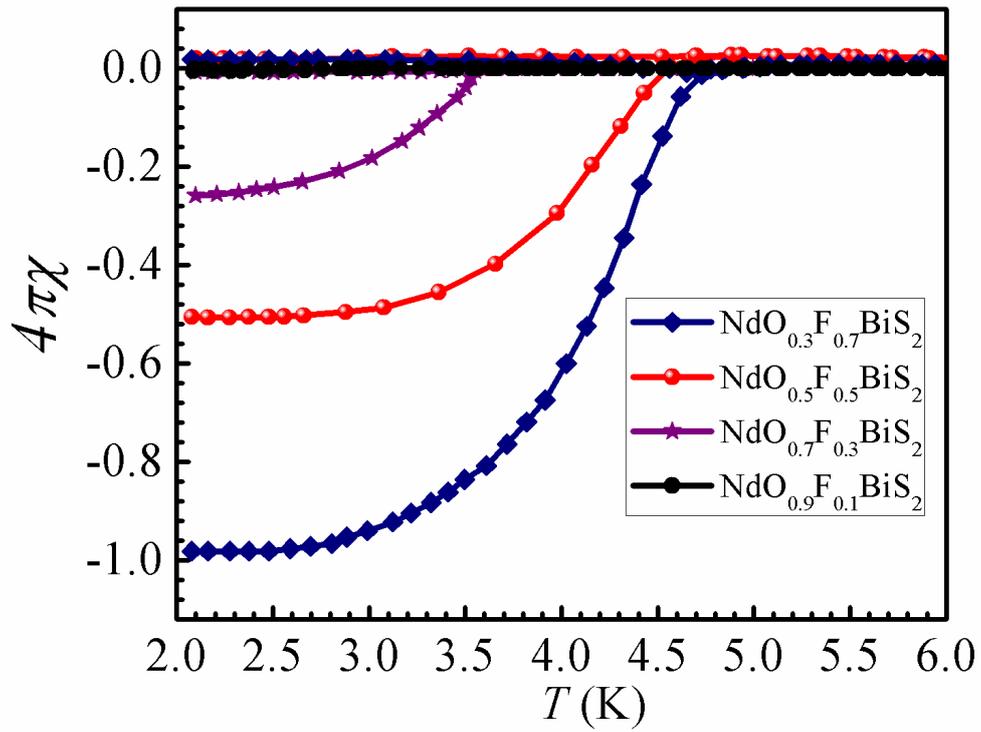

**Figure 3**

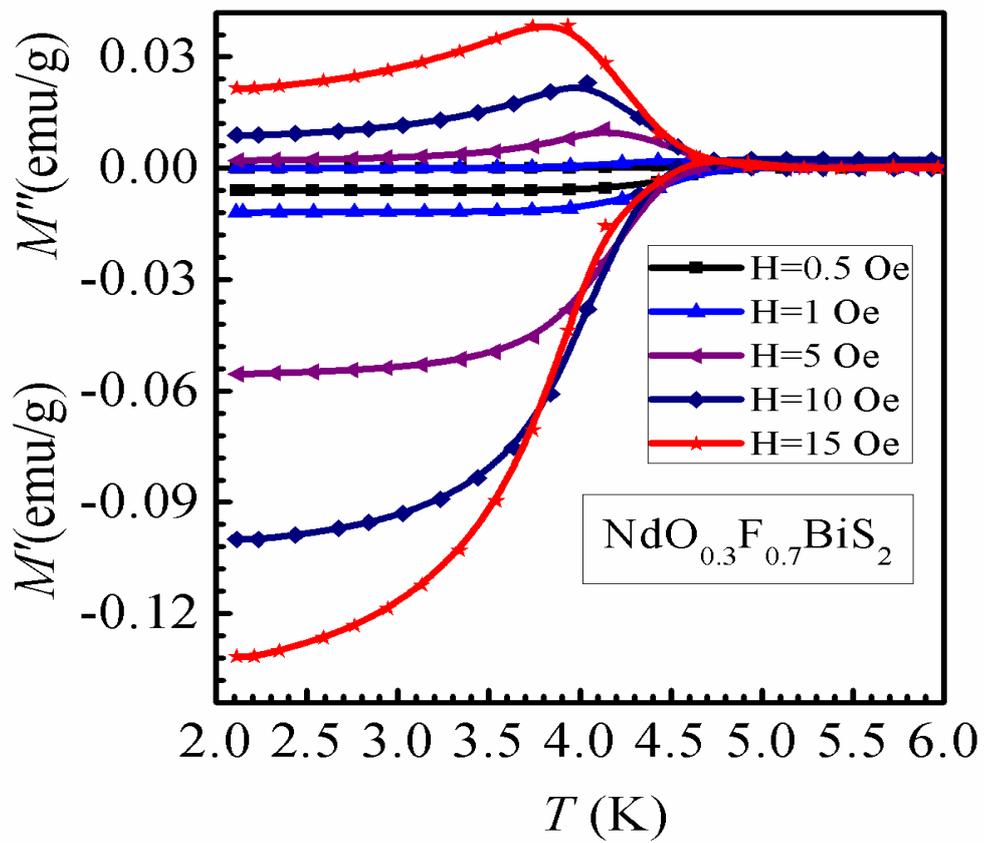

**Figure 4**

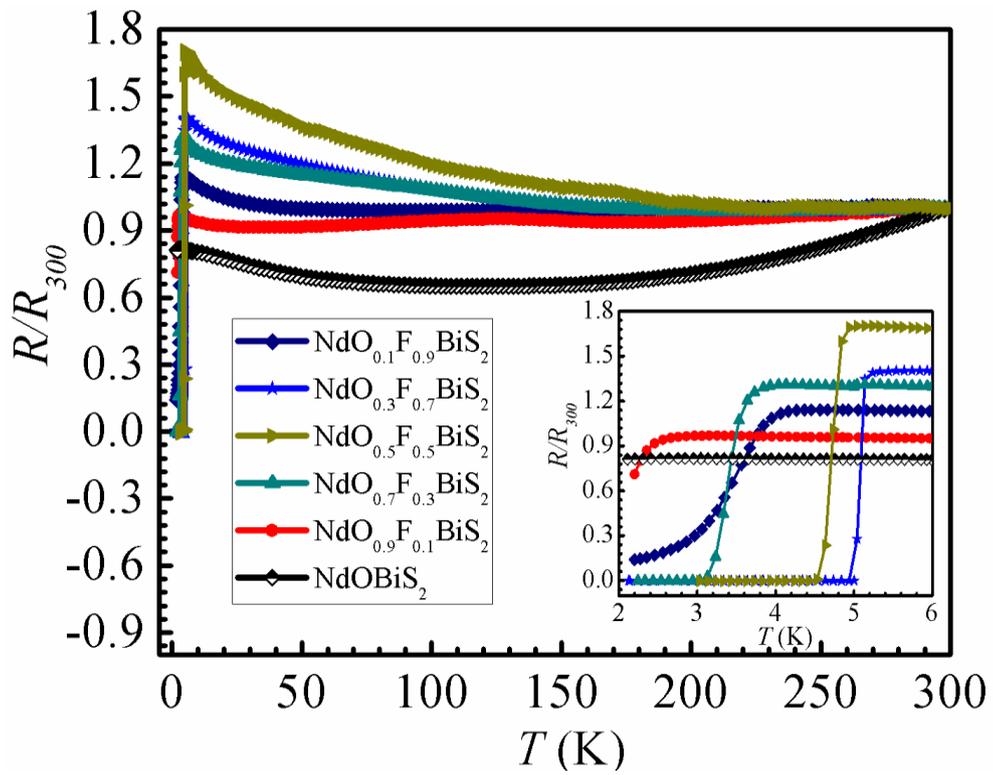

**Figure 5:**

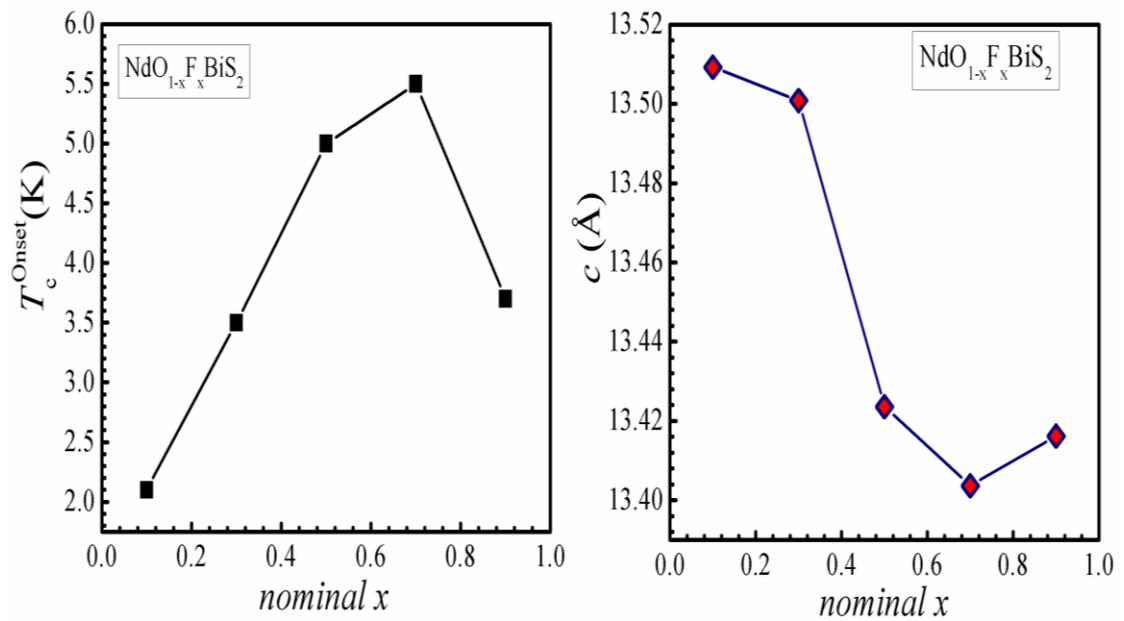

**Figure 6:**

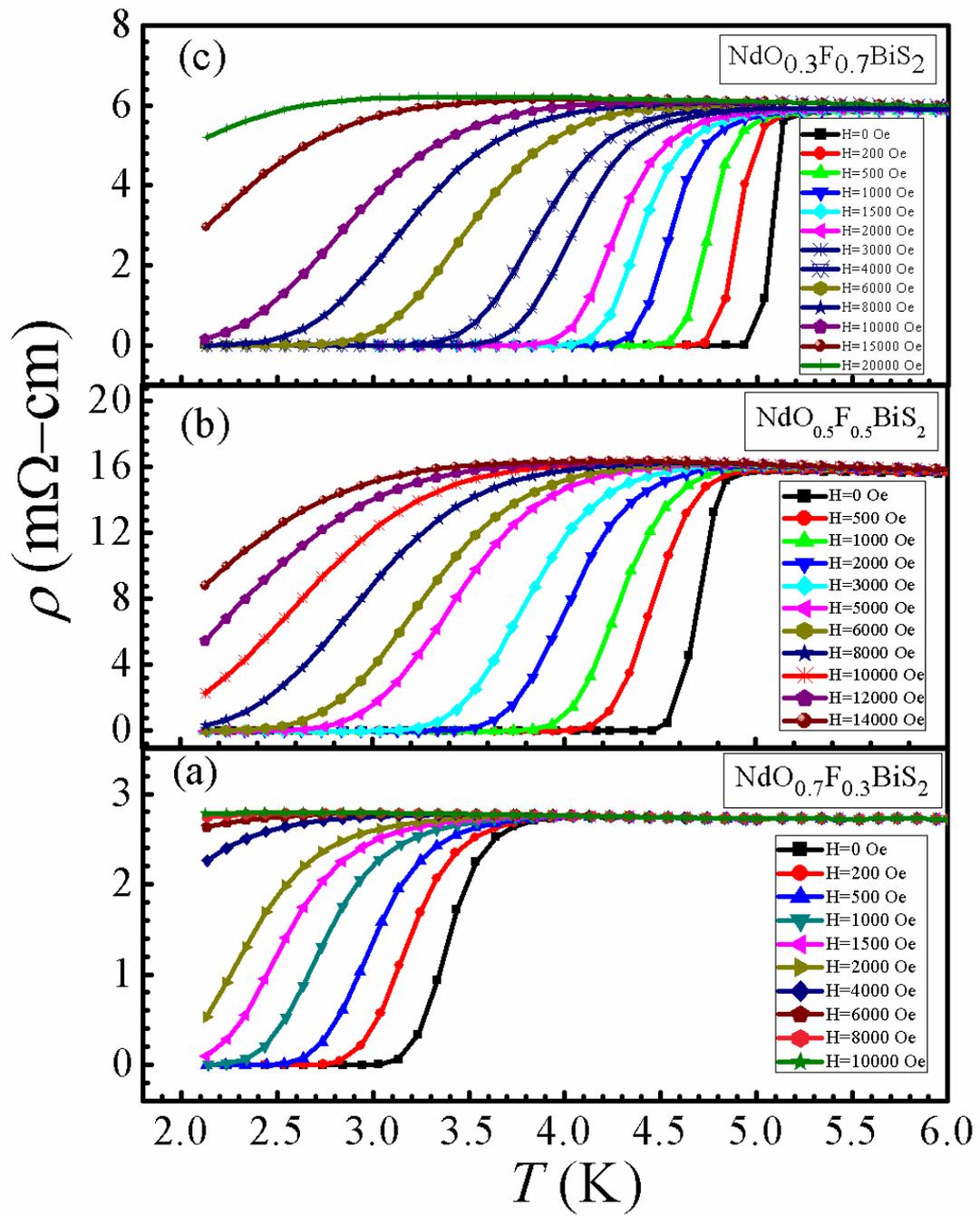

**Figure 7:**

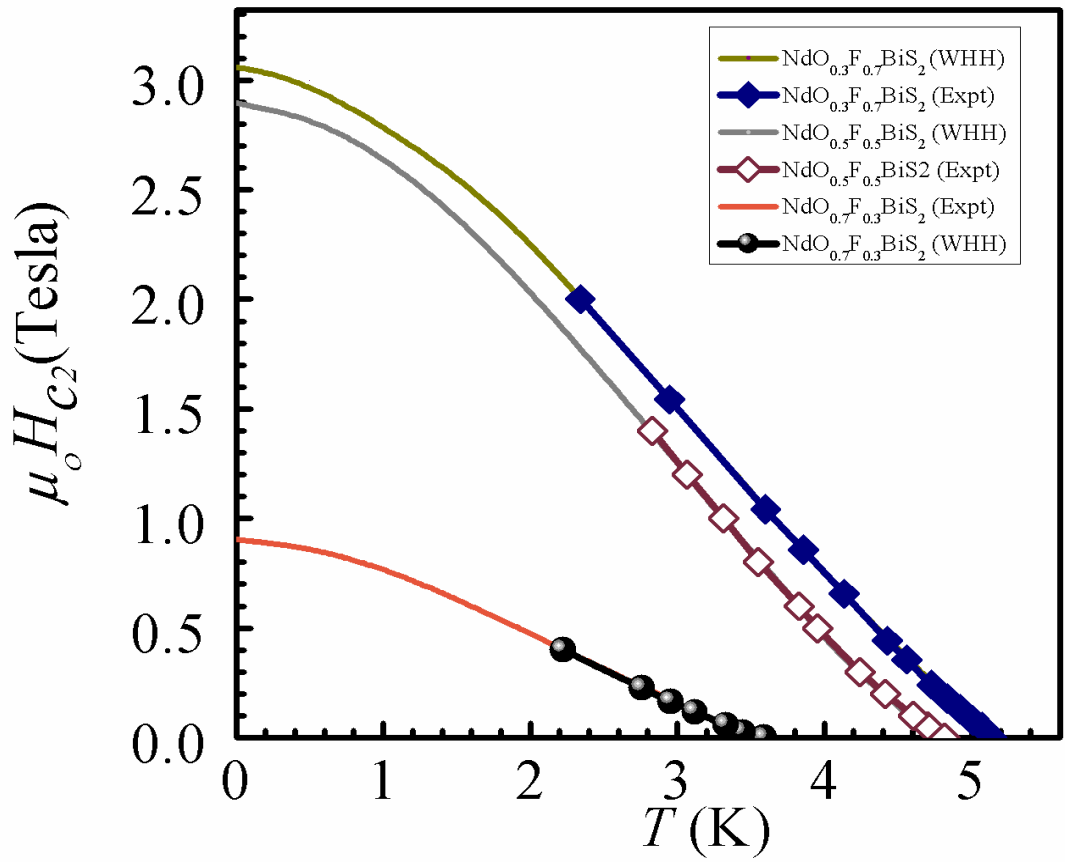

**Figure 8:**

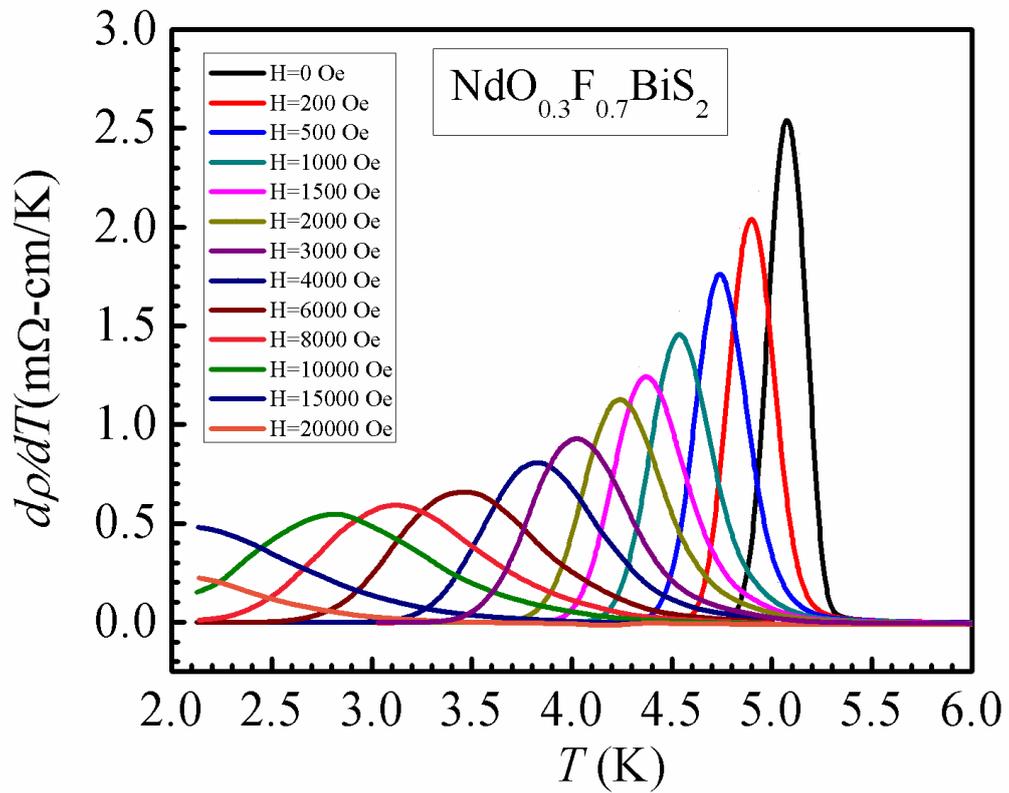

**Figure 9:**

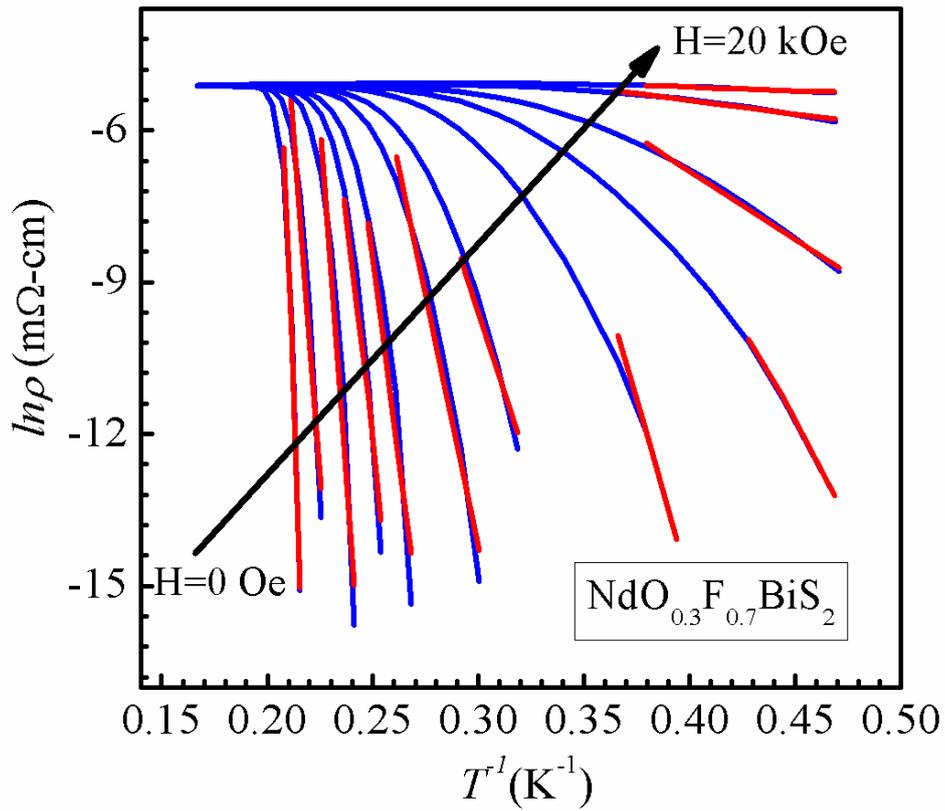

**Figure 10:**

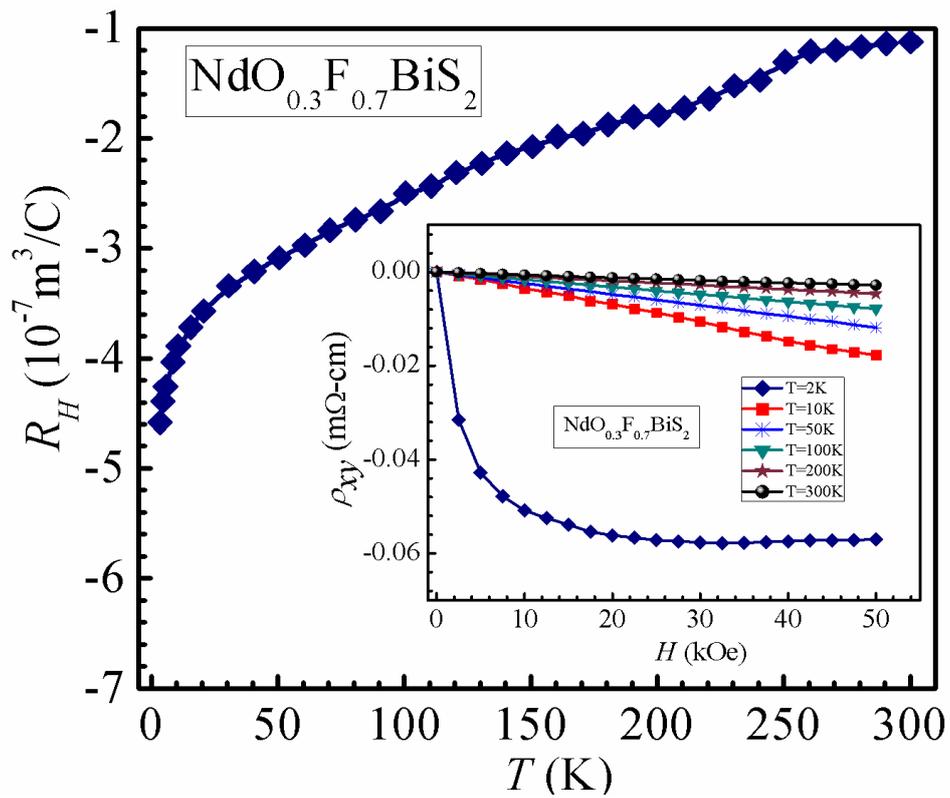